\documentclass[%
 reprint,
 article
 amsmath,amssymb,
 aps,
prb,
]{revtex4-1}
\usepackage{amsmath}
\usepackage{dcolumn}
\usepackage{graphicx}
\usepackage{float}

\begin{document}

\preprint{APS/123-QED}

\title{Synergy and competition between superconductivity and antiferromagnetism in FeSe under pressure }

\author{Guan-Yu Chen, Enyu Wang, Xiyu Zhu, and Hai-Hu Wen}\email{hhwen@nju.edu.cn}

\affiliation{National Laboratory of Solid State Microstructures and Department of Physics, Collaborative Innovation Center of Advanced Microstructures, Nanjing University, Nanjing 210093, China}

\begin{abstract}
Temperature dependence of resistivity under high pressures with magnetic fields parallel and perpendicular to the FeSe planes are measured in FeSe single crystals. It is found that the tetragonal to orthorhombic structural transition (nematic) temperature is suppressed by pressure and ends at around $P$ = 1.18 GPa. Below around 0.85 GPa, the superconducting transition shows a narrow width with no indication of antiferromagnetic order. While above this pressure, the superconducting transition temperature drops slightly forming a small dome of superconducting region with the maximum $T_c$ at around 0.825 GPa. Furthermore, just above this pressure, the superconducting transition exhibits an unusual large transition width which reaches about 6-8 K. This wide transition width is an intrinsic feature and does not change with magnetic field. In the high pressure region above 1.18 GPa, just accompanying the onset of superconducting transition, an upturn of resistivity immediately occurs, which is attributed to the formation of an antiferromagnetic order. This closely attached behavior of superconductivity and antiferromagnetic order indicates that these two orders have a synergy feature. The wide transition width in high pressure region is interpreted as the concurrence but spatially phase separated regions of the superconductivity and antiferromagnetic order. By applying a magnetic field, superconductivity is suppressed. Near the critical pressure 0.825 GPa and below, our data illustrate that an antiferromagnetic order emerges when superconductivity is suppressed. From the weak influence of magnetic field to the antiferromagnetic order, we conclude that it exists already below the small superconducting dome in the low pressure region. This shows a competing feature between superconductivity and antiferromagnetic order. Our results show duality features, namely synergy and competition between superconductivity and antiferromagnetic order under pressure in FeSe.

\begin{description}
\item[Subject Areas]Condensed Matter Physics, Superconductivity
\end{description}
\end{abstract}

\maketitle
\section{INTRODUCTION}
Since the discovery of superconductivity in La[O$_{1-x}$F$_x$]FeAs \cite{Hosono2008}, a lot of iron based superconductors (IBSs) have been discovered and synthesized, the highest superconducting transition temperature in bulk samples reaches about 55-57K \cite{RenZACPL,XuZAEPL,HHWen2009Europys}. It is found that the superconductivity occurs in the vicinity of antiferromagnetic order, together with the high values of superconducting critical temperatures, the IBS family has been regarded as another member of high-$T_c$ unconventional superconductors after the cuprate system\cite{1986cuperate}. Among all IBSs, the iron-selenium (FeSe), which has the simplest structure and shows $T_c$ = 8-10 K for bulk samples \cite{WuMKPNAS2008} and $T_c\geq$  65 K for one-unit-cell thick film \cite{FengDLWangYY}, stands out because of its rich and intriguing properties.

The unique character of FeSe is the complex relationship between nematicity, antiferromagnetism and superconductivity. In many systems of IBSs, there is an antiferromagnetic (AF) transition at $T_{AF}$ accompanying with the structural transition at $T_s$. The two transitions may follow up each other closely with the variation of temperature, for example in the 122 family\cite{Osborn2012PRB,Canfield2010PRB}. The optimal superconducting transition temperature appearers at the doping level where the AF order vanishes. These features suggest that AF spin fluctuation plays an important role in the pairing mechanism of IBSs. However, no AF transition is observed in FeSe above $T_c$ at ambient pressure, which makes it different from other IBSs. At ambient pressure, FeSe undergoes an structural transition from tetragonal to orthorhombic at around $T_s$ = 90 K \cite{Boehmer2013PRB}. The nematic state is established immediately around $T_s$. One of the possible pictures concerning the nematic transition is the lifting of the degeneracy of the $d_{xz}$ and $d_{yz}$ orbitals, which leads to the anisotropy of the in-plane electronic structure \cite{Shimojima2014PRB,Takahashi2014PRL}.

Furthermore, FeSe displays many interesting features under pressure. Recent investigations\cite{Kothapalli2016NC,J.G.Cheng2016NC,Terashima2016PRB} on FeSe single crystals illustrate an unique phase diagram concerning the systematic evolution with pressure. By increasing pressure, the structural transition temperature $T_s$ is suppressed and vanishes around a certain pressure, where a second transition at $T_{AF}$ = 20 K starts to emerge, which have been identified as an AF transition by nuclear magnetic resonance (NMR) \cite{WeiqiangYu2016PRL} and $\mu$SR \cite{Bendele2012muSR} measurements. Upon increasing pressure, the superconducting transition temperature $T_c$ increases initially, which is followed by a slight drop before increasing again. This small dome-like superconducting region shows a local maximum around 1.0 GPa. Above 1.5 GPa, both superconducting transition temperature $T_c$ and AF order transition temperature $T_{AF}$ increase simultaneously, which is supposed to be the evidence of cooperative relationship between superconductivity and antiferromagnetism. However, $T_c$ increases with the pressure from 2 GPa to 5 GPa, which forms a second dome of superconductivity, while $T_{AF}$ goes up with pressure and reaches 45 K around 5 GPa. At higher pressures, the AF order becomes unstable and superconducting transition temperature reaches a maximal value with $T_c$ = 38 K, which is nearly four times of that at ambient pressure. Finally, when the pressure is higher than about 8 GPa, the AF temperature $T_{AF}$ drops down following a dome like feature, while the superconducting transition temperature drops down with a much slower rate showing a separation of these two orders. The highest $T_c$ is realized where the AF order starts to collapse, which may indicate the competition between superconductivity and antiferromagnetism. However, a clear picture and fundamental reason for the evolution of $T_{AF}$ and $T_c$ are still unclear.

Concerning the superconducting gap structure and pairing mechanism, the FeSe system also exhibits a variety of intriguing physical properties and distinct features. The information about superconducting gaps is very important for determining the pairing mechanism of FeSe. Unfortunately, the exact structure of superconducting gaps is still under debate. A V-shaped spectrum seen by scanning tunneling microscopy (STM) suggests a nodal gap structure \cite{Q.K.Xue2011Science}. In contrast, results from transport measurements, such as specific heat measurements \cite{L.Jiao2016SR,hhwen2017PRB} and thermal conductivity measurement \cite{Tailleffer2016PRL}, favor a highly anisotropy but nodeless gap structure. A shoulder \cite{Y.Sun2017PRB} or a jump \cite{hhwen2017PRB} of specific heat coefficient around 1 K have been observed in some measurements. The STM measurements and related orbital selective analysis reveal the anisotropic but nodeless gap structure\cite{Davis2016Science}, which is supported by the theory of orbital-selective pairing model\cite{HirschfeldTheory}. Furthermore, the comparable values of superconducting gap $\Delta$ and Fermi energy $E_F$ put FeSe close to the Bardeen-Cooper-Schrieffer (BCS) to Bose-Einstein-condensation (BEC) crossover region, which has attracted tremendous attentions \cite{Matsuda2014PNAS,Matsuda2016NC}. All these unique features indicate that the FeSe is not only a very good system to investigate the pairing mechanism of superconductivity, but also the interplay between superconductivity and many intertwined orders.

In this paper, we report systematic resistivity measurement under finely tuned pressures up to 2.5 GPa in order to establish a detailed temperature-pressure phase diagram of FeSe. This phase diagram displays the evolution of $T_s$, $T_{AF}$ and $T_c$. We find a closely attached behavior between $T_{AF}$ and $T_c$ in the intermediate pressure region, which indicates that the superconductivity and antiferromagnetism have a synergy feature. In addition, for suppressing the superconductivity, we applied magnetic fields up to 16 T with directions both parallel to $c$-axis and $ab$-planes. This can help us to explore the possible existence of antiferromagnetic order under the superconducting dome. We track the evolution of $T_{AF}$ down to 0.825 GPa under the superconducting dome, this indicates that the AF order emerges when superconductivity is suppressed. This shows the competing feature between superconductivity and antiferromagnetism.

\section{EXPERIMENT}
The single crystals were grown using the chemical vapor transport technique \cite{hhwen2017PRB}. In order to have perfect homogeneity of samples, we use polycrystalline FeSe powder as the starting material. The FeSe powder was synthesized with solid state reaction method. The mixture of Fe and Se powders in a molar ratio of 1.04:1 was reacted at 850 $^\circ$C for 2 days followed by a quench at 400 $^\circ$C. Then the FeSe powder is mixed with KCl and AlCl$_3$ which were used as the transport agent \cite{Boehmer2013PRB}. The molar ratio is FeSe:KCl:AlCl$_3$ = 1:2:4 in the mixture, and they were put into the bottom of a quartz tube. All these procedures were carried out in a glove box filled with argon gas. After evacuating and sealing the quartz tube, it was placed into a horizonal tube furnace and heated up to 430 $^\circ$C to melt the transport agent. After 30 hours, the temperature of one end of the quartz tube that without reactant was changed to 370 $^\circ$C to create a temperature gradient, at the meantime the temperature of another terminal of the tube was kept at about 430 $^\circ$C. This growing status was kept for 6 weeks, and finally FeSe single crystals were obtained at the colder end of the quartz tube.

The FeSe crystals were characterized by measurements of magnetization and resistivity. The magnetization measurement was carried out by using a superconducting quantum interference device (SQUID-VSM 7 T, Quantum Design).  Two samples from the same batch were chosen for measurement of resistivity under pressure. The hydrostatic pressure was applied by a HPC-33 piston-type pressure cell which is provided by Quantum Design. The resistivity was measured in a physical property measurement system (PPMS 16 T, Quantum Design) by a standard four probe method. During the resistive measurements, magnetic fields up to 16 T were applied parallel to the $c$-axis and $ab$-plane of the crystals.

\section{RESULTS AND DISCUSSION}

\begin{figure}
\includegraphics[width=8.5cm]{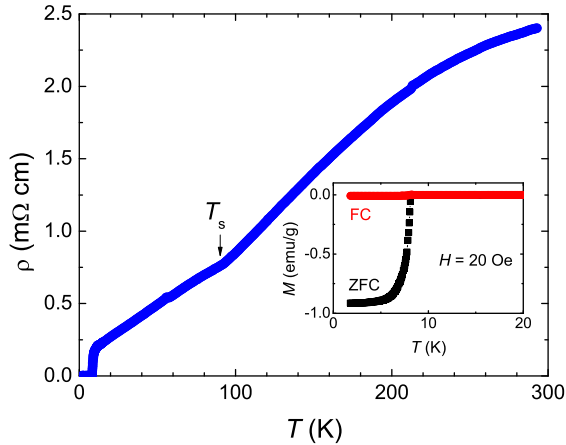}
\caption{\label{fig:1} Temperature dependence of resistivity for an FeSe single crystal sample at zero magnetic field at ambient pressure. The structural transition occurs at the temperature $T_s$ which is marked by the black arrow. The inset shows the magnetization measured by the ZFC and FC modes at a magnetic field of 20 Oe.}
\end{figure}

\subsection{Characterizations of samples}
The temperature dependence of resistivity $\rho(T)$ of one FeSe single crystal at zero magnetic field and ambient pressure is shown in Fig.~\ref{fig:1}. In the normal state, a metallic behavior can be observed. A kink that marked by a black arrow on the curve can be found clearly at around $T_s \approx$ 93 K. This anomaly is corresponding to the structural transition from tetragonal phase to orthorhombic phase, and also near the temperature that nematic state is established. Under a pressure the structural transition temperature will drop down. This structural transition has been proved by high-pressure x-ray diffraction (XRD) measurement \cite{Bohmer2016NC}. The onset of superconducting transition temperature at ambient pressure is $T_c^{onset} \approx$ 9.5 K, which is determined by the crossing point of the extrapolated lines of the normal state and the steep transition region. The sample enters zero resistivity state at $T_{c0} \approx$ 8.5 K. The superconducting transition width $\Delta$$T_c$ is about 1 K, which is defined by the relation $\Delta$$T_c$ = $T_c^{onset}$ - $T_{c0}$. By extrapolating the normal state resistivity curve down to 0 K, the value $\rho_n$(0 K) is estimated. Therefore, the residual resistivity ratio (RRR), which is determined by the ratio of $\rho_n$(300 K)/$\rho_n$(0 K), is about 19. The inset in Fig.~\ref{fig:1} shows the magnetization of the FeSe crystal measured in zero-field-cooled (ZFC) and field-cooled (FC) modes with the external magnetic field of 20 Oe. The superconducting transition temperature $T_c$ determined by the overlapped point of the ZFC curve and FC curve is about 8.4 K, which is quite close to the value of $T_{c0}$ determined by zero resistivity. The superconducting volume estimated from the temperature dependence of magnetization in the ZFC mode is larger than 100 \% because of the demagnetization effect, which indicates the bulk superconductivity of the FeSe crystal. The large RRR value, sharp superconducting transitions, and large superconducting volume confirm the high quality of our samples, which guarantees the reliability of further study.

\subsection{Temperature-pressure phase diagram at zero field}

\begin{figure}
\includegraphics[width=8.5cm]{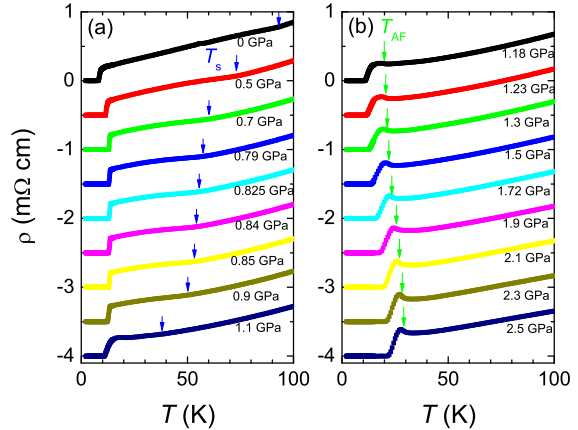}
\caption{\label{fig:2} Temperature dependence of resistivity of one FeSe single crystal at zero field under various pressures below 100 K. (a) Data measured under different pressures ranging from ambient pressure to 1.1 GPa. (b) Data measured under different pressures ranging from 1.18 to 2.5 GPa. The blue arrows in (a) mark the structural transitions at $T_s$. The AF transition temperatures at $T_{AF}$ are marked by green arrows in (b).}
\end{figure}

In order to establish the temperature-pressure phase diagram of FeSe, the resistivity at various pressures has been measured and the data are displayed in Fig.~\ref{fig:2}. The curves of resistivity versus temperature at different pressures are shifted by a finite offset along the vertical axis for clarity. The structural transition temperature $T_s$, which shows up as a kink in the curve of resistivity versus temperature, is marked by blue arrows. By applying pressure, $T_s$ is suppressed gradually and the kink feature of resistivity becomes gradually invisible. As the pressure goes up to 1.1 GPa, $T_s$ is suppressed down to below 40 K and vanishes at higher pressures. The superconducting transition temperature $T_c$ increases with pressure from the beginning, and drops slightly with further increasing pressure until it rises up again at around 1.1 GPa. The $T_{c0}$ reaches a maximum value around 0.825 GPa. At the pressure of about 1.18 GPa, just accompanying the onset of superconducting transition and the collapse of tetragonal phase, a second transition manifested by a remarkable upturn of resistivity starts to emerge at about 20 K. This feature is marked by the green arrows in Fig.~\ref{fig:2}(b). According to previous studies, this upturn of resistivity is corresponding to the AF order transition that has been proved in high-pressure NMR \cite{WeiqiangYu2016PRL} and $\mu$SR \cite{Bendele2012muSR} measurements. This has also been corroborated by the work of other groups, which reaches a consensus that this upturn of resistivity is closely related to the formation of the AF order \cite{J.G.Cheng2016NC,Terashima2016PRB}, therefore we determine the temperature of $T_{AF}$ by taking the point at which the derivative $d\rho/dT$ has the strongest negative slope. This upturn of resistivity appears on all $\rho(T)$ curves with pressures between 1.18 GPa and 2.5 GPa. As pressure goes up from 1.1 GPa, the $T_{AF}$ increases progressively and more pronounced feature of the resistivity upturn can be observed, which indicates that the AF order is stabilized under higher pressure. At 2.5 GPa which is the highest pressure achieved in our present study, $T_{AF}$ finally reaches about 29 K. In this procedure, the superconducting transition temperature $T_c$ also rises up together with the increase of $T_{AF}$. This indicates an intimate relationship between superconductivity and antiferromagnetism in FeSe. We must emphasize that the pressure for observing this upturn of resistivity seems to be sample dependent\cite{Ji-HoonKang2016SST}. In most samples, this upturn shows up at the similar pressure among different groups. However in some samples, it appears at a much higher pressure when the applied field is zero. We assume that this is related to the subtle change of the internal properties of the samples, for example the concentration of the interstitial irons. Further efforts are needed for clarifying this discrepancy.

\begin{figure}
\includegraphics[width=8.5cm]{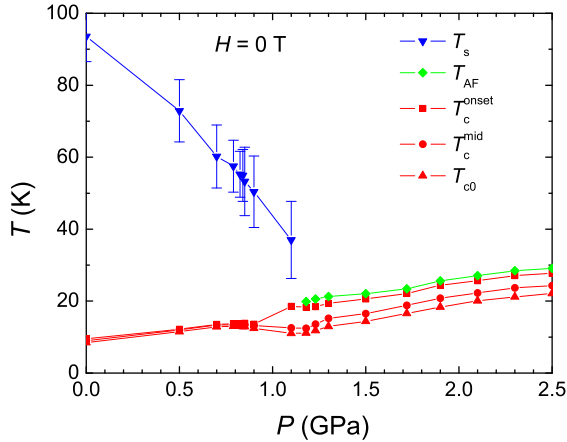}
\caption{\label{fig:3} The temperature-pressure phase diagram of an FeSe single crystal at zero field with the data extracted from Fig.2. The structural transition temperature $T_s$ is marked by the blue downward triangles. The green diamond symbols represent the AF transition temperature $T_{AF}$. The superconducting temperatures defined by three criterions $T_c^{onset}$, $T_c^{mid}$ and $T_{c0}$ are plotted as functions of pressure, which are represented by red squares, circles, and triangles, respectively.}
\end{figure}

To unravel the relationship between nematicity, antiferromagnetism and superconductivity, we extract the values of these characteristic temperatures from the data shown in Fig.~\ref{fig:2} and present them in Fig.~\ref{fig:3}. This figure also serves as a temperature-pressure phase diagram of FeSe, in which the evolution of $T_s$, $T_{AF}$ and $T_c$ as a function of pressure can be clearly recognized. The structural transition temperature $T_s$ and AF order temperature $T_{AF}$ are determined by the anomalies in resistivity as stated above. What surprises us is that, when the pressure is increased to the region beyond about 0.9 GPa, the superconducting transition becomes very wide and the width does not change too much across the whole pressure region up to 2.5 GPa. This feature seems to be intrinsic and one can also find the same phenomenon from previously published data \cite{Ji-HoonKang2016SST}. In order to investigate the pressure dependent transition width, we determine three different transition temperatures $T_c^{onset}$, $T_c^{mid}$ and $T_{c0}$ which are displayed in Fig.~\ref{fig:3}. The onset transition temperature $T_c^{onset}$, which features that the FeSe crystal starts to enter superconducting state, is determined by two different methods depending on the pressure regions. Below 1.18 GPa, where the antiferromagnetic order does not show up yet, the resistivity curve in normal state shows a metallic behavior and the superconducting transition is quite sharp, $T_c^{onset}$ can be defined as the crossing point of the extrapolated lines of the normal state and the steep transition part. Above 1.18 GPa, where the AF order is observed, an upturn signature of resistivity starts to emerge and $T_c^{onset}$ is defined as the peak position of resistivity. Below the temperature corresponding to this peak, resistivity drops down quickly and smoothly without any other features, therefore it is reasonable to define this peak position as the first occurrence of superconductivity. The middle transition temperature $T_c^{mid}$ is determined from the point with the half resistivity of the peak value, namely by $\rho$($T_c^{mid}$) = $\rho$($T_c^{onset}$)/2. The zero resistivity is realized at $T_{c0}$, which shows the formation of well connected or full volume of superconducting state. The correlations of the three superconducting transition temperatures are plotted together with those corresponding to the nematicity and AF transitions in Fig.~\ref{fig:3}. As already known, there is no any feature of AF order at ambient pressure above $T_c$, and the structural transition occurs at around $T_s \approx$ 93 K. The system enters a nematic state but with absence of the AF order, which is still puzzling. However, a specific heat anomaly around 1.08 K was reported in a similar crystal \cite{hhwen2017PRB}, which might be the signature of AF transition under the superconducting state. This needs of course more experimental verifications. With the increase of pressure, $T_s$ decreases monotonously and becomes hardly definable around 1.18 GPa, where the AF order starts to emerge at $T_{AF} \approx$ 20 K. Although the signature of $T_s$ can not be observed in the resistivity measurement above this threshold pressure, it does not mean the absence of the structural transition. The results from recent high-pressure x-ray diffraction measurement \cite{Bohmer2016NC} shows that the orthorhombic (nematic) distortion will still take place at $T_{AF}$, which is displayed as a more profound split of the structural parameters along $a$-axis and $b$-axis. The coexistence of nematicity and antiferromagnetism might indicate the close relationship between them. In this pressure region, the three superconducting transition temperatures move up with pressure initially and form a small superconducting dome with the maximum value, typically the $T_{c0}$, around 0.825 GPa. The dome is followed by a slight drop of $T_{c0}$, which leads to a $T_{c0}$ valley around 1.1 GPa, and the AF order appears just from this pressure. According to previous high-pressure Hall effect study \cite{J.G.Cheng2017PRL}, in the pressure range where the AF order is induced, the Fermi surface of FeSe will undergo a reconstruction, in which the charge carriers change from electron-type to dominantly hole-type. The presence of hole pocket is helpful to the stabilization of AF order due to the Fermi surface nesting mechanism. However, the carrier density will be suppressed by such Fermi surface reconstruction, which leads to the competition between superconductivity and AF order. At higher pressures, both $T_{AF}$ and three $T_c$'s values just increase in a parallel way, and the onset of superconducting transition temperature $T_c^{onset}$ are always attached to $T_{AF}$, which indicates that the Cooper pairing occur in accompany with the formation of AF order. The closely attached behavior of superconductivity and the AF order indicates that these two orders have a synergy feature.

\begin{figure}
\includegraphics[width=8.5cm]{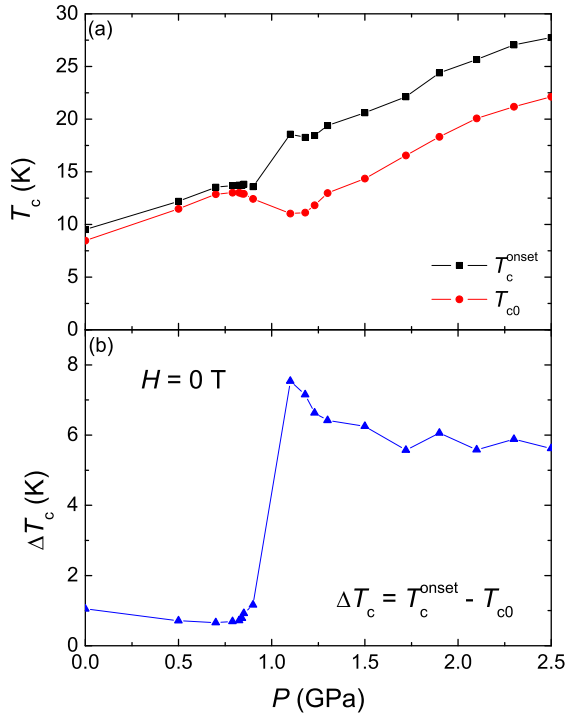}
\caption{\label{fig:4} (a) Pressure dependence of two superconducting transition temperatures $T_c^{onset}$ and $T_{c0}$, which are represented by black squares and red circles. (b) Pressure dependence of superconducting transition width $\Delta$$T_c$, which is calculated by the difference between $T_c^{onset}$ and $T_{c0}$.}
\end{figure}

In the pressure range where the antiferromagnetic transition sets in, a remarkable change of the superconducting transition width $\Delta$$T_c$ can be observed in both of Fig.~\ref{fig:2} and Fig.~\ref{fig:3}. The evolution of $T_c^{onset}$ and $T_{c0}$ as function of pressure are plotted in Fig.~\ref{fig:4}(a), in this way the transition width can be analyzed in a more detailed way. The superconducting transition width $\Delta$$T_c$ is determined by the difference between the zero and onset transition temperatures, namely $\Delta$$T_c$ = $T_c^{onset}$ - $T_{c0}$, and the result is shown in Fig.~\ref{fig:4}(b). As one can see, the superconducting transition shows a narrow width below 0.9 GPa, where the AF order is absent. Surprisingly, just above this pressure, the superconducting transition exhibits an unusual large transition width which reaches about 6 - 8 K. The remarkable change of the transition width occurs where the antiferromagnetic order sets in, which indicates the competition feature between superconductivity and antiferromagnetism.

\subsection{The emergence of antiferromagnetic order below the superconducting transition at low pressures}
Although the temperature-pressure phase diagram shown in Fig.~\ref{fig:3} provides important message for understanding the physics in FeSe, it remains however a puzzling whether the AF order exists at ambient and low pressures. In the phase diagram reported by previous studies \cite{Bendele2012muSR,Bohmer2016NC,Ji-HoonKang2016SST}, the AF order was not observed above the superconducting transition temperature $T_c$, while it is difficult to know whether this order appears below $T_c$ since most of the features of the AF order would be shielded by superconductivity. Thus in most of these studies, the extending of AF order under superconducting dome from the high pressure side is either stopped at a certain pressure or was plotted as a dashed line entering the superconducting dome based on speculations. Neutron scattering experiment finds that below $T_c$ there are strong and unique AF spin fluctuations \cite{ZhaoJunNatMat}. Some of the specific heat measurements also reveal that there is an anomaly at around 1.08 K, which was argued to be the possible existence of the AF order stabilized probably by the interstitial irons\cite{hhwen2017PRB}. However some other experiments show the absence of this specific heat anomaly \cite{Hardy}. Therefore it is quite crucial to know whether there is an AF order hidden below the superconducting state. One way to trace out this AF order is through the AF phase line discovered above the critical pressure, for example about 1.18 GPa in our experiment. We can use magnetic field to suppress the superconductivity and to see whether this AF order will show up. In order to explore the possible hidden AF order in FeSe below the superconducting transition temperature, we measured resistivity under magnetic fields up to 16 T with the field parallel to $c$-axis and $ab$-planes. We show clear evidence of the emergence of the AF order below $T_c$. The results are detailed below.

\begin{figure}
\includegraphics[width=8.5cm]{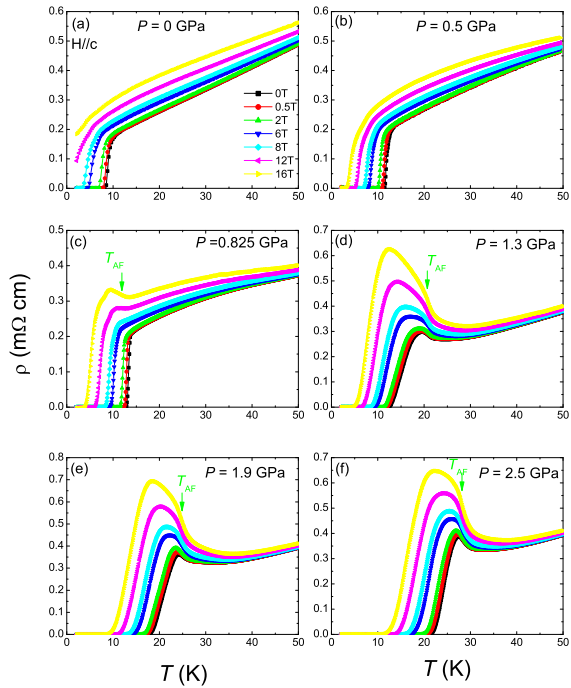}
\caption{\label{fig:5} Temperature dependent resistivity under pressures for the FeSe single crystal (sample1) in different magnetic fields $H$ = 0 T, 0.5 T, 2 T, 6 T, 8 T, 12 T, 16 T, the magnetic field is applied parallel to $c$-axis of the crystal. The measurements were done under many different pressures, but the data are selectively shown for 0 GPa in (a), 0.5 GPa in (b), 0.825 GPa in (c), 1.3 GPa in (d), 1.9 GPa in (e) and 2.5 GPa in (f), respectively. The green arrows represent the AF transition at $T_{AF}$, which are manifested by the upturn behavior of the curve of resistivity versus temperature.}
\end{figure}

The temperature dependencies of resistivity measured under selected pressures of 0, 0.5, 0.825, 1.3, 1.9 and 2.5 GPa and different external magnetic fields are shown in Fig.~\ref{fig:5} with the field parallel to $c$-axis of the FeSe crystal. At ambient pressure, it can be seen that $T_c$ decreases with increasing magnetic field, and the state with zero resistivity is not observed above 2 K and 12 T. The resistivity curve in the normal state shows a clear enhancement compared with that in zero field, and a bending down occurs even at a high temperature under magnetic fields. Although the data exhibit obvious magnetoresistance, the giant magnetoresistance that appears as an extreme upturn behavior in magnetic field reported in previous study \cite{Matsuda2014PNAS,Matsuda2016NC} is not observed in our present sample. The difference between the magnitude and features of magnetoresistance may be caused by the slightly different doping levels or impurity concentrations of the FeSe samples. The giant magnetoresistance may be induced by the semi-metal behaviors of FeSe in which the electron and hole bands are both very shallow. The properties of FeSe are sensitively influenced by the stoichiometric ratio of elements, such as the different density of interstitial iron atoms. At 0.5 GPa, the superconductivity is getting enhanced under pressure, which is even retained up to 16 T and the resistivity curve is very rounded in the normal state. The situation changes when the pressure reaches 0.825 GPa. There is no any signature of AF order at this pressure and zero field, which is evidenced by the monotonic temperature dependence of resistivity above $T_{c0}$. However, with increasing magnetic field, the superconductivity is suppressed, an upturn of resistivity gradually emerges. As shown in Fig.~\ref{fig:5}(c), a shoulder appears on the $\rho(T)$ curve at temperatures above $T_c$ when a 12 T magnetic field is applied, which further evolves into a peak at 16 T. The shape of the peak appearing at 16 T is similar to the upturn signature that occurring at zero field when the pressure is higher, as shown in Fig.~\ref{fig:2} and Fig.~\ref{fig:5} (d)-(f). We presumably conclude that the resistivity peaks have the same origin, namely they indicate the formation of an AF order. It is worth to mention that the $T_{AF}$ obtained by the point corresponding to the rapid uprising of resistivity at 16 T under 0.825 GPa is lower than the $T_{c0}$ measured at zero field under same pressure, which means that the trace of AF order hidden in the superconducting state have been exposed. With the pressure going up to 1.3 GPa, the peak appears even at zero field and it is getting enhanced by magnetic field. A remarkable upturn behavior at 16 T can be observed and the value of resistivity at $T_c^{onset}$ is almost two times higher compared with that measured at zero field. Interestingly, the value of $T_{AF}$ shows null or weak field dependence, which manifests that the AF order is quite robust against magnetic field. At higher pressures, it can be seen in Fig.~\ref{fig:5}(e) and (f) that the shape of the resistivity peak above $T_c$ is similar to the one shown in Fig.~\ref{fig:5}(d) and $T_{AF}$ keeps growing up progressively with pressure. From the systematic evolution of the resistivity peaks at different pressures and magnetic fields we feel more confident that the upturning of resistivity induced by applying a high magnetic field at low pressures, e.g. 0.825 GPa is corresponding to the emergence of the AF order. It is important to note that $T_{AF}$ is field independent, which indicates that the occurrence of the AF order is only dependent on pressure, but the AF order is not induced by magnetic field. Another evidence is that the superconducting transition width $\Delta$$T_c$ exhibits an unusual change with the appearance of AF order when the field is zero. As shown in Fig.~\ref{fig:4}(a), the critical pressure is around 0.9 GPa. Below this pressure the resistive transition width is rather narrow at zero field. However, when the resistivity peak appears under a high magnetic field, the transition width gets immediately broadened. This supports the picture again that the newly emergent resistivity peak under magnetic field at the pressure of 0.825 GPa is corresponding to the AF order which is otherwise hidden by the superconductivity. Since the superconductivity occurs in a close link with the AF order, we would speculate that in the temperature region associated with the broad transition width, i.e., between $T_c^{onset}$ and $T_{c0}$, there might be a phase separation of the superconducting phase and the AF order. This looks like the intermediate state of a type-I superconductor with the spatial separation of the superconducting region and the normal state region.

\begin{figure}
\includegraphics[width=8.5cm]{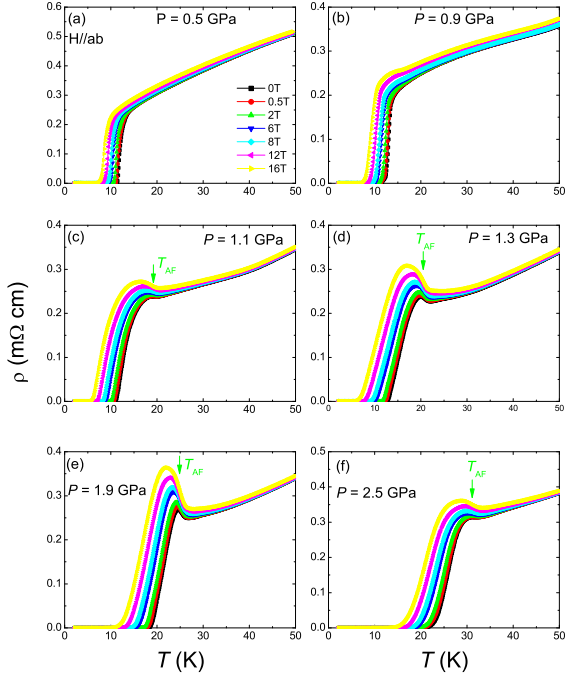}
\caption{\label{fig:6} Temperature dependent resistivity under pressures for FeSe single crystal (sample2) at different magnetic fields $H$ = 0 T, 0.5 T, 2 T, 6 T, 8 T, 12 T, 16 T, with magnetic field parallel to $ab$-plane of the crystal. The data are selectively shown for 0.5 GPa in (a), 0.9 GPa in (b), 1.1 GPa in (c), 1.3 GPa in (d), 1.9 GPa in (e) and 2.5 GPa in (f), respectively. The green arrows represent the antimagnetic transition at $T_{AF}$, which are determined by the rapid upturn behavior in the resistivity curves.}
\end{figure}

For comparison, the resistivity with magnetic fields parallel to $ab$-plane of FeSe crystal is also measured, and the data are selectively shown in Fig.~\ref{fig:6} with pressures of 0.5, 0.9, 1.1, 1.3, 1.9 and 2.5 GPa. The magnitude of magnetoresistance effect is quite different for $H\parallel$$ab$ and $H\parallel$$c$. For $H\parallel$$c$, the resistivity of FeSe in normal state is sensitive to external magnetic field, while it shows much weaker field dependence for $H\parallel$$ab$. This anisotropic property is related to the quasi-two-dimensional feature of the Fermi surface. Just like many other IBSs with layered-structure, the upper critical field along $ab$-plane is larger than that along $c$-axis, which leads to difficulty in suppressing superconductivity and the resistivity peak shown in Fig.~\ref{fig:5}(c) at P = 0.825 GPa becomes hardly visible for $H\parallel$$ab$. For example, at 0.9 GPa, where the upturn behavior has appeared in resistivity curve for $H\parallel$$c$ under a field of 16 T, but it only shows a little enhancement of resistivity without the peak under a magnetic field at 16 T. It seems that the resistivity peak or the AF order would appear when the field is higher than 16 T. A shoulder of resistivity eventually appears at 1.1 GP at zero field and it develops into a strong peak gradually with increasing magnetic field. The magnitude of this resistivity peak for $H\parallel$$ab$ is much smaller than that for $H\parallel$$c$, but the values of $T_{AF}$ at 1.1 GPa determined by the resistivity measured with $H\parallel$$ab$ and $H\parallel$$c$ are almost identical, indicating again that the $T_{AF}$ is nearly field independent. As in the case of H||c-axis, by applying higher pressures, more pronounced peaks and resistivity upturns are induced and the peak with the strongest amplitude appears at 1.9 GPa.

\begin{figure}
\includegraphics[width=8.5cm]{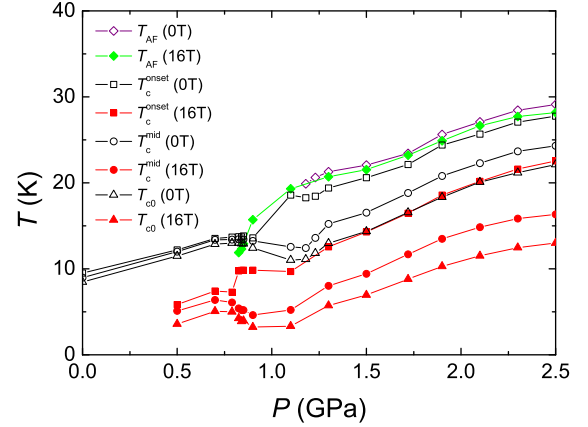}
\caption{\label{fig:7} Evolution of the antiferromagnetic transition temperature $T_{AF}$ determined from the resistivity upturn and superconducting transition temperature $T_c$ as functions of pressure. The data at zero field and 16 T are put together for comparison, which are represented by open and filled symbols, respectively.}
\end{figure}

Since the superconducting state is more easily suppressed for magnetic field parallel to $c$-axis, we present in Fig.~\ref{fig:7} the temperature-pressure phase diagram of FeSe at 16 T for $H\parallel$$c$. The transition temperatures at 16 T are summarized from the resistivity data shown in Fig.~\ref{fig:5}, and the transition temperatures at zero field, which have been displayed in Fig.~\ref{fig:3}, are plotted together for comparison. For clarity, the transition temperatures at 16 T and 0 T are represented by filled and open symbols, respectively, and the definition of the transition temperatures at 16 T is the same as that for 0 T. The structural transition temperature is not shown here, since $T_s$ keeps nearly unchanged with magnetic field. The superconductivity is completely suppressed by magnetic field at ambient pressure and $T_c$ rises slowly with increasing pressure. Although the $T_c$ at 16 T is shifted down compared with that at 0 T, the evolution of three different $T_c$s as a function of pressure displays similar feature. At 16 T, $T_c$ forms a small superconducting dome around 0.7 GPa, which is followed by a valley with a minimum around 0.9 GPa before rising up again with increasing pressure. The similar features of $T_c$ have been observed at 0 T as well, and the only difference is that the small superconducting dome and valley at 16 T are shifted to lower pressure region compared with that at 0 T. At zero external field, the signature of AF is poorly defined below 1.18 GPa. However, it is found that $T_{AF}$ at 16 T at low pressures is smoothly connected to the data of $T_{AF}$ at 0 T above 1.18 GPa, and its trace is extended down to 0.825 GPa with the help of applying a high magnetic field. In the low pressure region, the slope of $T_{AF}$ versus pressure changes abruptly and a rapid entering into the superconducting dome at 0 T can be observed. The emergence of the AF order under a high magnetic field in the superconducting dome confirms the coexistence of AF order and superconductivity in this region. Thus, we can conclude that the AF order might exist at pressures below 0.825 GPa and even extend to ambient pressure, which is covered by the superconducting state. More efforts are needed to clarify this picture. Unfortunately the highest magnetic field applied in our experiment is only 16 T, it would be interesting to measure the resistivity at ambient or low pressures with higher magnetic fields. If we have a closer look at the pressure dependence of $T_{AF}$ and $T_c$, it seems that the curve of $T_{AF}(P)$ intercepts the small superconducting dome (specially $T_{c0}$) at round 0.825 GPa where there is a maximum or optimal superconducting transition temperature. This inspires us to assume that there might be some kind of quantum critical behavior at this low pressure point. Our data can only give a preliminary hint, which deserves to be carried out by further experimental efforts.

\section{SUMMARY}
In summary, we have studied the evolution of nematicity, antiferromagnetism and superconductivity in FeSe crystals by measuring the in-plane resistivity under pressures up to 2.5 GPa and magnetic fields up to 16 T in parallel and perpendicular directions of the FeSe-plane. At low pressures, where the antiferromagnetic order is absent above $T_c$, the nematicity temperature drops down while the superconducting transition temperature rises up upon applying a pressure, indicating a competing feature between them. In this low pressure region with the absence of the AF order, the superconducting transition width is rather narrow. With the pressure going up, $T_{c0}$ drops clearly and the signature of $T_s$ is almost vanishing above 1.1 GPa. Just accompanying the drop of $T_{c0}$, the transition width exhibits a unusual large value $\Delta$$T_c$ that reaches 6 - 8 K at higher pressures. Meanwhile, the antiferromagnetic order starts to emerge at $T_{AF} \approx$ 20 K when the pressure is further increased. The drop of $T_{c0}$ accompanied by the occurrence of the AF order suggest that the superconductivity competes with the AF order. However, at higher pressures, the onset superconducting transition is closely attached to the occurrence of the AF order, this shows a synergy behavior between them. At pressures slightly below the critical value, by applying magnetic field up to 16 T, the signature of AF order covered by superconductivity is recovered again. The AF order is thus extended to 0.825 GPa by applying a high magnetic field. Our data show that the AF order enters into the superconducting dome in the low pressure region, which indicates that antiferromagnetic order may already exist in superconducting state.

\begin{acknowledgments}
We thank Christoph Meingast for helpful discussions. This work was supported by the National Key Research and Development Program of China (2016YFA0300401,2016YFA0401700), and the National Natural Science Foundation of China (NSFC) with the projects: A0402/11534005, A0402/11374144.
\end{acknowledgments}

\end{document}